# 90-DEGREE BLOCH DOMAIN WALL STRUCTURE IN A CUBIC CRYSTAL WITH A NEGATIVE MAGNETIC ANISOTROPY

## Svitlana A. Dyachenko, Bogdan M. Tanygin, Olexandr V. Tychko

*National Taras Shevchenko University of Kyiv, RadioPhysics Faculty, prosp. Acad. Glushkova, 2, building 5, pasat@univ.kiev.ua*

*An influence of strong sample demagnetization field on the structure of plane 90-degree Bloch domain walls in a cubic (001)-crystal with a negative first constant of magnetic anisotropy is considered.*

1. For some samples (thin films, magnetic particles, etc.) their demagnetization field becomes the factor influencing a structure of a domain wall [1]. It may be connected with a deviation of a magnetization vector **M** from medium easy magnetization axes at presence of enough strong sample demagnetization fields. In particular, strong demagnetization field in thin epitaxial magnetic ferrite-garnet (001)-films results in a reorientation of the **M** in domain volumes from easy magnetization axes to the film plane [2].

At homogeneous magnetization distribution, the volume orientation of the **M** in the thin plate is determined by minimum of energy density $e_A = e_M + e_{MA}$, where $e_{MA}$ and $e_M$ - volume energy density of the magnetic anisotropy and demagnetization field $\mathbf{H_d}$ respectively [1]: $e_M = -(\mathbf{MH_d})/2 = 2\pi M^2 \alpha_3^2$; $e_{MA} = K_1(\alpha_1^2\alpha_2^2 + \alpha_2^2\alpha_3^2 + \alpha_1^2\alpha_3^2) + ..$, where $M$ — saturation magnetization (**M**=M**б**, **б** - unit vector); **б** $=(\alpha_1, \alpha_2, \alpha_3)$; $\alpha_1$, $\alpha_2$ and $\alpha_3$ — directing cosines in coordinate system $Oxyz$ with axes along <100>-, <010>- and <001>- direction accordingly; $K_1$ — first magnetic anisotropy constant. For (001)-plate at $p = 2\pi M^2/|K_1| \geq 0.5$ and $K_1 < 0$, the minimum density value of the energy $e_A(\alpha_1, \alpha_2, \alpha_3)$ that is equal $e_A^\infty(\alpha_1, \alpha_2, \alpha_3) = K_1/4$, is achieved at $\alpha_3^\infty = 0$, $\alpha_1^\infty = \alpha_2^\infty = \pm 1/\sqrt{2}$. The values $\alpha_i^\infty$ ($i=1,2,3$) set possible directions of the domain magnetization vectors $\mathbf{M_i}$ and domain wall type. These directions coincide with intermediate magnetization axes (crystallographic directions such as <110>) in (001)-plane of the cubic crystal with $K_1 < 0$. In that case for isolated Bloch domain wall that divides two domains with magnetization vectors $\mathbf{M_1}$ and $\mathbf{M_2}$ ($\mathbf{M_1} = M\mathbf{m_1}$, $\mathbf{M_2} = M\mathbf{m_2}$, $\mathbf{m_1}$ and $\mathbf{m_2}$ - unit vectors specifying limiting directions of **M** in the domain volumes) the angle $2\alpha$ between vectors $\mathbf{m_1}$ and $\mathbf{m_2}$ determines domain wall type ($2\alpha$-degree) [3]: $\alpha = arccos[1+(\mathbf{m_1 m_2})/2]^{1/2}$. Hence possible domain wall types are 90-degree and 180-degree. Thus a 90-degree domain wall (DW) in a cubic crystal with $K_1 < 0$ appears at $p \geq 0.5$. The structure of this DW is described hereinafter.

2. If the **M** direction describes by polar $\theta$ and azimuth $\varphi$ angles that are accordingly counted out from a normal **n** to DW plane and a vector $\Delta\mathbf{M} = \mathbf{M_2} - \mathbf{M_1}$, then the **M** spatial distribution in the DW volume is described by variable $\varphi$ ($\theta$ does not change in DW volume owing to preservation of normal components of **M** for plane Bloch DW: $cos\theta = cos\alpha \, cos\lambda$). Here $\lambda$ is an angle between the normal **n** and a plane of vectors $\mathbf{m_1}$ and $\mathbf{m_2}$ that coinciding with a plane of a plate, describes rotation **n** around of the vector $\Delta\mathbf{M}$ and sets of the DW plane orientation. Orientation of $\mathbf{m_1}$ and $\mathbf{m_2}$ is set by the values $\varphi$ that equal accordingly $\varphi_1$ and $\varphi_2$. For isolated plane DW the specific energy $\sigma$ can be presented in a traditional form [1]:

$$\sigma = 2\int_{\varphi_1}^{\varphi_2} \sqrt{A sin^2\theta \cdot (e_A(\theta,\varphi) - e_A^\infty(\theta,\varphi_1))} \, d\varphi,$$

where $A$ — an exchange constant, $\varphi_1 = -\varphi_0 < \varphi < \varphi_2 = +\varphi_0$ and $\varphi_1 = \varphi_0 < \varphi < \varphi_2 = 2\pi - \varphi_0$ - accordingly for the right- and left-handed rotation of **M**, $\varphi_0 = arccos(sin\lambda/\sqrt{2-cos^2\lambda})$ — half of the angle between $\mathbf{m_1}$ and $\mathbf{m_2}$ projections on DW plane ($\alpha < \varphi_0 < \pi - \alpha$) [3]. In the chosen coordinate system connected with the DW plane the energy density $e_A$ can be presented in the following form:

$$e_A(\theta,\varphi) = p\{3+(\cos2\theta-\cos2\lambda)(1-\cos2\varphi)+\cos2\varphi-\cos2\theta\cos2\lambda(3+\cos2\varphi)-4\cos\varphi\sin2\theta\sin2\lambda\}/8+$$
$$+\{97-12(\cos2\theta+\cos2\varphi)-9\cos4\lambda+4\cos2\varphi(3\cos4\lambda-4\cos2\theta)+\cos4\varphi(28\cos2\theta-21-3\cos4\lambda)-$$
$$-4\cos2\theta\cos4\lambda(5-4\cos2\varphi-\cos4\varphi)- \cos4\theta[7(3-4\cos2\varphi+\cos4\varphi)+\cos4\lambda(35+28\cos2\varphi+\cos4\varphi)]\}/512+$$
$$+16\sin2\theta\sin4\lambda(\cos3\varphi-\cos\varphi)-8\sin4\theta\sin4\lambda(7\cos\varphi+\cos3\varphi)$$

3. Orientation dependencies of specific energy $\sigma/\sin\psi$ are presented Fig.a. Here $\psi$ is an angle between (001)- and DW planes. DW parameters (for equilibrium orientation) dependencies on sample demagnetization field are resulted in Fig.b.

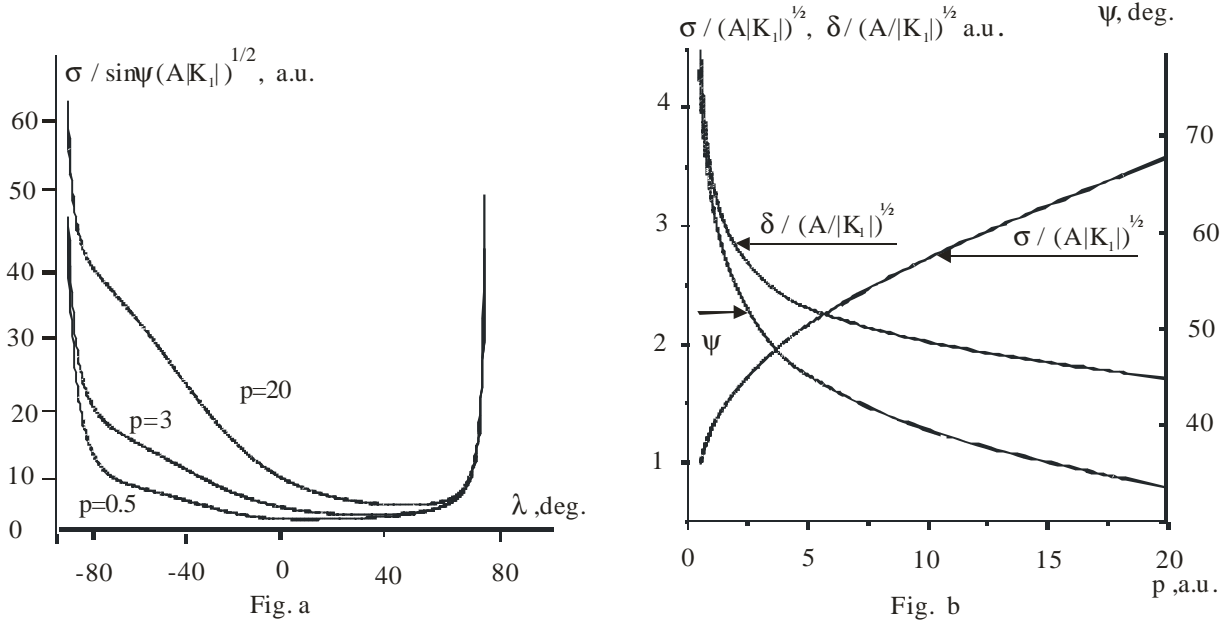

Fig. a　　　　　　　　　　　　Fig. b

For $p \geq 0.5$ possible equilibrium DW orientations are set by the expression:

$$p = [23-30\cos2\lambda-37\cos4\lambda+39\cos6\lambda+13\cos8\lambda-9\cos10\lambda+\cos12\lambda+$$
$$+\sqrt{2(3-\cos2\lambda)}(49\sin\lambda+3\sin3\lambda+24\sin5\lambda-\sin7\lambda-6\sin9\lambda+\sin11\lambda)]/$$
$$\{4[25+21\cos2\lambda-10\cos4\lambda-5\cos6\lambda+\cos8\lambda-\sqrt{2(3-\cos2\lambda)}(9\sin\lambda+12\sin3\lambda+2\sin5\lambda-\sin7\lambda)]\}.$$

DW have equilibrium orientations in a range $13.8^0 \leq \lambda \leq 90^0$. With $p$ growth the general tendency of specific energy $\sigma/\sin\psi$ increase is remained for DW.

Depending on DW orientation their thickness $\delta$ [4] is determined by the following expression:

$$\delta=8\sqrt{A/|K_1|}\left(\arccos\left(\frac{\sqrt{2}\sin\lambda}{\sqrt{3-\cos2\lambda}}\right)\sqrt{3-\cos2\lambda}\right)(9+8p(3+2\cos2\lambda-\cos4\lambda)-4\cos2\lambda-2\cos4\lambda -$$
$$-4\cos6\lambda+\cos8\lambda+\sqrt{2(5+4\cos2\lambda-\cos4\lambda)}(\sin2\lambda-8p\sin2\lambda-4\sin4\lambda+2\sin6\lambda))^{\frac{1}{2}}$$

General tendency of DW thickness decrease is kept with growth $p$ for examined DW.

1. A. Hubert, R. Shafer Magnetic domains. The analysis of magnetic microstructures. Berlin: Springer-Verlag, 1998.
2. Sohatsky V., Kovalenko V. //Journal De Physique IV. Colloque CI. Suppl. Journ. De Physique III.-1996.-V.7,№3.-P.C1-699 - C1-702.
3. O.A. Antonyuk, A.V. Tychko, V.F. Kovalenko // Alloys and Compounds. 2004. 369. 112-116.
4. B.A. Lilley //Phill. Mag. 1950. 41. 319. 792-813.